\documentstyle[12pt]{article}
\begin{document}

\topmargin 0pt

\oddsidemargin -3.5mm

\headheight 0pt

\topskip 0mm
\addtolength{\baselineskip}{0.20\baselineskip}
\begin{flushright}
MIT-CTP-$2702$
\end{flushright}
\begin{flushright}
SOGANG-HEP $228/97$
\end{flushright}
\vspace{0.5cm}
\begin{center}
    {\large \bf  New Gauge Invariant Formulation of the Chern-Simons Gauge 
Theory}
\end{center}
\vspace{0.5cm}

\begin{center}
 Mu-In Park
%\footnote{Electronic address: mipark@physics.sogang.ac.kr}
$ ^{a}$ and Young-Jai Park
%\footnote{Electronic address: yjpark@ccs.sogang.ac.kr} 
$ ^{b}$\\
{$^{a}$ Center for Theoretical Physics, Massachusetts Institute of
 Technology, \\ 
Cambridge, MA 02139 U.S.A.\\
$^b$ Department of Physics, Sogang University, \\
 C.P.O. Box 1142, Seoul 100-611, Korea} \\
\end{center}
\vspace{0.5cm}
\begin{center}
    {\bf ABSTRACT}
\end{center}
A new gauge invariant formulation of the relativistic scalar field interacting
with Chern-Simons gauge fields is considered. This formulation is consistent  
with the gauge fixed formulation. Furthermore we find that canonical (Noether)
Poincar\'e generators are not gauge invariant even on the constraints surface 
and do not satisfy the (classical) Poincar\'e algebra. It is the
 improved generators, constructed from the  
symmetric energy-momentum tensor, which are (manifestly) gauge
 invariant and obey the  
classical Poincar\'e algebra.
\vspace{3cm}
\vspace{2.5cm}
\begin{flushleft}
%January 1997 \\
\end{flushleft}

\newpage

Recently gauge invariant pertubative analysis using Dirac's dressed matter 
fields [1] has been of considerable interest in QED and QCD, especially in 
relation to the infrared divergence and quark confinement problems [2]. In 
general, there are two approaches in quantum field theory: the gauge invariant
formulation (GIF) and the gauge fixed formulation (GFF). The latter is the 
conventional one, where one chooses a gauge; in the former on the other hand, 
one does not fix the gauge but works with gauge invariant quantities. However 
as for the formalism itself, it is not clear how the results in GIF can be 
matched to 
GFF even though this matching is considered in several recent analyzes [3].

Furthermore, a similar gauge independent 
Hamiltonian analysis [5] $\acute{ a}~ la $ Dirac [1, 4] has been recently 
considered for the Chern-Simons (CS) gauge theory with matter fields [5]. 
Actually after the CS gauge theory was invented, there arose several debates 
about the gauge dependence of the spin and statistics transmutation phenomena 
for the charged matter fields, since the analysis was carried out with 
specific gauge fixing [6-8]. So, with the formulation without gauge fixing, 
one can expect to resolve this debate since one is not confined to a specific 
gauge. But the result of the recent gauge independent analysis for this 
problem in Ref. [5] is questionable since there is no room for spin 
transmutation. This is in sharp contrast to the well known spin transmutation 
of GFF [6-8]. 

In this paper we shall provide a new gauge invariant Hamiltonian formulation 
which is consistent with GFF. By introducing a physically plausible 
assumption, we find a new set of equations for Dirac's dressing function 
$c_{k}({\bf x},{\bf y})$. Furthermore, we provide, for the first time, a 
simple interpretation of how the dressing $c_{k}({\bf x},{\bf y})$ is related 
to the gauge fixing and how GIF is matched to GFF. As a byproduct we find that
canonical (Noether) Poincar\'e generators are not gauge invariant
``even on the constraints surface'' and do not 
satisfy the (classical) Poincar\'e algebra. It is the improved
generators, constructed  
from the symmetric (Belinfante) energy-momentum tensor [9], which are
(manifestly) gauge  
invariant and obey the classical Poincar\'e algebra.
This effect is essentially due to the CS term and is important for genuine 
spin transmutation in the relativistic CS gauge theory. Furthermore the fact 
that only the symmetric energy-momentum tensor, not the canonical one, is 
meaningful is consistent with Einstein's theory of gravity. All
results in this paper are at the classical level, but not at the
quantum level. 

Our model is the Abelian CS gauge theory with massive relativistic complex 
scalars [5, 6]
\begin{eqnarray}
{\cal L}=\frac{\kappa}{2} \epsilon^{\mu \nu \rho}A_{\mu}\partial_{\nu}A_{\rho}
+(D_{\mu}\phi)^{*}(D^{\mu}\phi)-m^{2} \phi^{*} \phi,
\end{eqnarray}
where $\epsilon^{012}=1$, $g_{\mu \nu}$=diag(1,--1,--1), and $D_{\mu}=
\partial_{\mu}+iA_{\mu}$. ${\cal L}$ is invariant up to the total divergence 
under the gauge transformations
$\phi \rightarrow exp[-i\Lambda] \phi,~~A_{\mu} \rightarrow A_{\mu} +
\partial_{\mu}\Lambda, $
where $\Lambda$ is a well-behaved function such that
$\epsilon^{\mu \nu \lambda}
\partial_{\mu} \partial_{\nu} \Lambda=0$. By using the basic (equal time)
 bracket (called Faddeev-Jackiw (FJ) bracket [10])
\begin{eqnarray}
&&\{ A^i({\bf x}), A^j({\bf y}) \} =\frac{1}{\kappa} \epsilon^{ij}
 \delta^{2} ({\bf x}-{\bf y}), \nonumber \\
&&\{ \phi({\bf x}), \pi ({\bf y}) \}=
 \{\phi^{*} ({\bf x}), \pi^{*}({\bf y}) \} = 
\delta^{2}({\bf x}-{\bf y}), ~~\mbox{others~vanish}
\end{eqnarray}
with $\pi=(D_{0}\phi)^{*}$, $\pi^{*} =D_{0} \phi$, there remains the 
constraint 
$T \equiv J_0-\kappa B \approx 0$ ${\acute a}~la$ FJ, where $J_0$ is
the time component of the conserved matter current  
$J_{\mu}=i [ (D_{\mu} \phi)^{*} \phi -\phi^{*} D_{\mu} \phi]$ and
$B= \epsilon_{ij}\partial_{i} A^{j}$ is the magnetic field . Here, we note 
that in the FJ bracket method, $T \approx 0$ is the only constraint
and the primary  
constraints of the Dirac bracket method, $\pi_{0}
\approx 0,~\pi_i -\frac{\kappa}{2} \epsilon_{ij} A^{j}\approx 0$, need not be 
introduced.

Now in order to develop the manifestly gauge invariant Hamiltonian formulation
we introduce the gauge invariant matter and gauge fields
\begin{eqnarray}
\hat{\phi}({\bf x}) \equiv \phi({\bf x}) exp(iW),~
\hat{\pi}({\bf x}) \equiv \pi({\bf x}) exp(-iW),~
{\cal A}_{\mu}({\bf x})\equiv{A}_{\mu}({\bf x})-\partial_{\mu}W,
\end{eqnarray}
and their complex conjugates with $W=\int c_{k}({\bf x},
 {\bf z}) A^{k}({\bf z}) d^{2} z$. The Dirac dressing function 
$c_{k}({\bf x},{\bf z})$ satisfies
\begin{eqnarray}
\partial^{k}_{z} c_{k}({\bf x},{\bf z})=-\delta^2({\bf x}-{\bf z}). 
\end{eqnarray}
Here, we note that there are infinitely many solutions of $c_k$ which 
satisfy (4) and the gauge invariance of fields in (3) should be
understood on each solution hypersurface but not on the entire
solution space. 
Furthermore, it should be noted that at the quantum level, due to the
commutation relation 
$[{A^i}_{op} ({\bf y}), \hat{\phi}_{op}({\bf x}) ]=\frac{\hbar}{\kappa}
\hat{\phi}_{op}({\bf x})\epsilon_{ik} c_{k} ({\bf x}, {\bf y})$,
the gauge invariant operator $\hat{\phi}_{op}({\bf x})$ creates one charged
$(~[J_0 ({\bf y}), \hat{\phi}_{op}({\bf x}) ]=\delta({\bf x}-{\bf y}) 
\hat{\phi}_{op}({\bf x}) ~)$ particle together with the gauge varying vector 
field ${{ a}^{i}}({\bf y})=\frac{\hbar}{\kappa}\epsilon_{ik} c_{k} ({\bf x}, 
{\bf y})$, as well as the gauge invariant point magnetic field ${b}({\bf y})
(=\epsilon_{ij}\partial_i a^j)=\frac{1}{\kappa}\delta^2({\bf y}-{\bf x})$.
[This situation is in contrast to the QED case where 
$\hat{\phi}_{op}({\bf x})$ 
creates the gauge invariant (physical) electron together with only the gauge 
invariant electric field [1, 2].] Now, returning to the classical level, with 
(3),
the Poincar\'e 
generators which being (manifestly) gauge invariant and satisfying the
Poincar\'e algebra become
\begin{eqnarray}
&&{P}^{0}_{s} = \int d^{2}x \left[ |{\hat{\pi}}|^{2}
      +|{\cal D}^{i} {\hat{\phi} } |^{2} +m^{2} |{\hat{\phi}}|^{2}
       \right], \nonumber \\
&&{P}^{i}_{s} = \int d^{2}x \left[{\hat{ \pi}}{\cal D}^{i}{\hat{\phi}} 
      +({\cal D}^{i}{\hat{\phi}})^{*} {\hat{\pi}^{*}}
       \right], \nonumber \\
&&{M}^{12}_{s} = \int d^{2}x~ \epsilon_{ij}x^{i} \left[
  {\hat{ \pi}}{\cal D}^{j}{\hat{\phi}} 
  +({\cal D}^{j} {\hat{\phi}})^{*}  {\hat{\pi}^{*}}
   \right], \nonumber \\
&&{M}^{0i}_{s} = x^{0} {P}^{i}_{s}-
 \int d^{2}x~ x^{i}\left[|{\hat{ \pi}}|^{2} 
 +|{\cal D}^{j}{\hat{\phi}}|^{2} +m^{2} |{\hat{\phi}}|^2 \right],
\end{eqnarray}
which are expressed only by the gauge invariant fields and ${\cal D}_i 
\equiv {\partial}_{i} +i {\cal A}_i$. These are $improved$ generators 
following the terminology of Callan $et~ al.$ [11] constructed from the 
symmetric (Belinfante) energy-momentum tensor [9]. Note that as far as we are 
interested in the dynamics of the physically relevant fields of (3), there are
no additional terms proportional to constraints in the Poincar\'e generators 
of (5) [4, 5]. 
Next, let us consider the transformations generated by Poincar\'e generators 
(5) for the gauge invariant quantities of (3). First of all, we consider the 
spatial translation generated by
$\{ {\hat{\phi}}({\bf x}),{P}^{j}_{s}\}=
\partial^j {\hat{\phi}}({\bf x}) -i {\hat{\phi}}({\bf x})
\int d^2 z \left(\partial^j_z c_k ({\bf x},{\bf z}) + \partial^j_x c_k
({\bf x},{\bf z})\right) {A}^k ({\bf z})$,
where we have dropped the terms 
$\int d^2 z \partial^i_z [c_k({\bf x},{\bf z}) 
{A}^k({\bf z})]$ and $
\int d^2 z \partial^k_z \left[c_k({\bf x},{\bf z}) 
{A}^j({\bf z})\right]$, which vanish  for sufficiently rapidly decreasing 
integrand.
This shows the translational anomaly ( following the terminology of Hagen 
$et~al.$ [6-8], $``$anomaly" means an unconventional contribution ). However, 
we assume 
that this anomaly should not appear in order that  
 ${\hat{\phi}}$ respond conventionally to transaltions.
This assumption is motivated by the fact that usual local fields have no 
translational anomaly, regardless of their spin or other properties. With 
this assumption, we obtain the condition that
$c_{k}({\bf x}, {\bf z})$ be translationally invariant
$\partial^i_z c_k ({\bf x},{\bf z})=
-\partial^i_x c_k ({\bf x},{\bf z})$, 
i.e., $c_k({\bf x},{\bf z})=c_k({\bf x}-{\bf z})$.
Furthermore, this condition also guarantees the correct spatial
translation law for all  
other gauge invariant fields in (3):
$\{ {{{\cal  F}}_{\alpha}}({\bf x}), { P}^j_{s} \}= \partial^j
{{{\cal  F}}_{\alpha}}({\bf x}) ,~ {{\cal  F}}_{\alpha}= ({{\cal A}_{\mu}},
{\hat{\phi}},{{\hat{\phi}}^*})$.

By applying similar assumption to the time translation we obtain
$\{ {{{\cal  F}}_{\alpha}}({\bf x}), { P}^0_{s} \}= \partial^0
{{{\cal F}}_{\alpha}}({\bf x})$
and using $
\int d^2 z \partial^k_z \left[c_k({\bf x}-{\bf z}) 
{ A}^0({\bf z})\right] = 0$
we further obtain the condition that $c({\bf x}-{\bf z})$ be time independent.
However, for the rotation and Lorentz boost, the anomaly is present since in 
that case it represents the spin or other properties of 
${\cal F}_{\alpha}$. The bracket with the Lorentz generator is expressed as
\begin{eqnarray}
\{ {\cal F}_{\alpha}({\bf x}), M^{\mu \nu}_s \}&=& 
   x^{\mu} \partial^{\nu}{\cal F}_{\alpha}({\bf x})
   -x^{\nu} \partial^{\mu} {\cal F}_{\alpha}({\bf x}) 
   +\Sigma^{\mu \nu}_{\alpha \beta} {\cal F}_{\beta}({\bf x})
   +\Omega^{\mu \nu}_{\alpha}({\bf x}), \nonumber \\
\Omega^{\mu \nu}_{\hat{\phi} }({\bf x})&=&
   -i {\Xi}^{\mu \nu} ({\bf x}) 
\hat{\phi} ({\bf x}), ~
\Omega^{\mu \nu}_{ \hat{\phi}^{*} }({\bf x})=
   i {\Xi}^{\mu \nu} ({\bf x}) 
\hat{\phi}^* ({\bf x}), \nonumber \\
\Omega^{\mu \nu}_{{\cal A}_{\beta}}({\bf x})&=&
  \partial_{\beta} {\Xi}^{\mu \nu} ({\bf x}),
\end{eqnarray}
where $\Xi ^{\mu \nu} =- \Xi^{\nu \mu}$,
${\Xi}^{12}({\bf x})= \epsilon_{ij} x_{i} {\cal A}^{j} ({\bf x})
+ ({1}/{\kappa})
 \int d^2z z_{k} c_{k}({\bf x}-{\bf z}) J_0({\bf z}),~ 
{\Xi}^{0i}({\bf x})=-x_i {\cal A}^0({\bf x})
-({1}/{\kappa}) \int d^2 z z_i 
  \epsilon _{kj} c_{k}({\bf x}-{\bf z}) J^{j}({\bf z})$.The anomalous
term $\Omega^{\mu \nu}_{\alpha}$ is gauge invariant.  
At first, it seems odd that the gauge invariant quantities do have the anomaly,
but as will be clear later, these quantities are nothing but the Hagen's 
rotational anomaly term and other gauge restoring terms in GFF. Before
establishing  
this, it is interesting to note that ${\cal A}_{\mu}$ can be re-expressed 
completely by the matter currents as 
${\cal A}_{i} ({\bf x}) \approx -({1}/{\kappa}) \int d^2 z \epsilon _{ik}
 c_{k} ({\bf x}-{\bf z}) J^0 ({\bf z}),~ {\cal A}_{0} ({\bf x}) = 
-({1}/{\kappa}) \int d^2 z \epsilon _{kj} c_{k} 
({\bf x}-{\bf z}) J^j ({\bf z})$ 
using the constraint $T \approx 0$ and the Euler-Lagrange equation of (1), 
$ \epsilon_{kj} J^j =F^{0k}$, respectively. These solutions are similar to the
Coulomb gauge solution [6] and hence imply the similarity of GIF to GFF with 
the
Coulomb gauge. However it should be noted that the Lorentz anomaly does not 
occur in the transformation of the current $J^{\mu}$, 
even though ${\cal A}_{\mu}$, which is expressed by $J^{\mu}$ as given
above, does  
have the anomaly.

Now, let us consider the basic brackets between the gauge invariant fields 
\begin{eqnarray}
\{ {\cal A}_i({\bf x}),{\cal A}_j({\bf y}) \} &=&
  \frac{1}{\kappa} \left[\epsilon_{ij}\delta^2({\bf x}-{\bf y})
  +\xi_{ij}({\bf x}-{\bf y}) + \partial_i^x \partial_j^y 
\Delta({\bf x}-{\bf y}) \right], \nonumber \\
\{ {\hat{\phi}}({\bf x}),{\hat{\phi}}({\bf y}) \} &=&
  -{\hat{\phi}}({\bf x}){\hat{\phi}}({\bf y}) \frac{1}{\kappa}
  \Delta({\bf x}-{\bf y}), \nonumber \\
\{ {\hat{\phi}}({\bf x}),{\hat{\phi}}^{*}({\bf y}) \} &=&
  -{\hat{\phi}}({\bf x}){\hat{\phi}}^* ({\bf y}) \frac{1}{\kappa}
  \Delta({\bf x}-{\bf y}), \nonumber \\
\{{\cal A}_i({\bf x}), \hat{\phi}({\bf y}) \} &=&
  -\frac{i}{\kappa}{\hat{\phi}}({\bf y}) \left[\epsilon_{ik}c_k({\bf y}-
  {\bf x})
   + \partial^x_i \Delta({\bf x}-{\bf y}) \right]. 
\end{eqnarray}
Here $\Delta({\bf x}-{\bf y})=\int d^2z \epsilon^{kj} c_k({\bf x}-{\bf z}) 
 c_j({\bf y}-{\bf z})$ and 
$\xi_{ij}({\bf x}-{\bf y})=\epsilon_{ik}\partial^y_j
c_k({\bf y}-{\bf x}) + \epsilon_{kj}\partial^x_i c_k({\bf x}-{\bf y})$. These 
results, together with the fact that corresponding quantum operator 
$\hat{\phi}_{op}({\bf x})$ creates the charged scalar particle together with 
the point magnetic flux at the point ${\bf x}$, seem to show resemblance of 
$\hat{\phi}$ to the 
$anyon~ field$ [12] but
it is found that this is not the case [6, 13].

Next, we consider how the gauge invariant results are matched to 
gauge fixed results. Actually this is connected with the gauge 
independence of the Poincar\'e algebra. The master formula for the matching
is, as can be easily proved
\begin{eqnarray}
\{ L_a,~ L_b \} \approx \{ L_a ,~ L_{b} \} _{D_{\Gamma}}.
\end{eqnarray}
Here $L_a$ is any gauge invariant quantity, where bracket with the first class
constraints $T$ of (4) vanishes, $ \{L_a,T\} \approx 0$. The left-hand side of
the formula (8) is the basic bracket of $L_a$'s. The right-hand side of (8)
is the Dirac bracket with gauge fixing function $\Gamma=0$, $det
|\{\Gamma, T\} | \neq 0$.
Moreover, in the latter case, since $\Gamma=0$ can be strongly implemented, 
$L_a$ can be replaced by 
${L_a}|_{\Gamma}$ that represents the projection of $L_a$ onto the surface 
${\Gamma}=0$. (This formula is implicit already in the recently developed 
Batalin-Fradkin-Tyutin formalism [14].)
Moreover, the left-hand side is gauge independent by construction since 
$L_a$'s and the the basic bracket algebra (2) are introduced gauge 
independently.
On the other hand, the Dirac bracket [4] depends explicitly on the chosen 
gauge $\Gamma$ in general. But there is one exceptional case, i.e., when the 
Dirac bracket is considered for the gauge invariant quantities. Our master 
formula (8) explicitly show this exceptional case: The Dirac bracket for the 
gauge 
invariant quantities $L_a$ or their projection ${L_a}|_{\Gamma}$ on the 
surface $\Gamma=0$ are still gauge invariant and equal to the basic bracket 
for the corresponding quantities.
Another important thing for the matching is to know how the defining equation 
(4) for $c_{k}({\bf x}-{\bf y})$ is modified in GFF. By considering 
$\hat{\phi}$ in a specific gauge and the residual gauge transformation of 
$\phi$ and 
$A_i$, one can find modified (but  still make $\hat{\phi}$ be gauge invariant)
 equation for $c_k({\bf x}-{\bf y})$. We provide here three typical cases:
$\mbox{a)~Coulomb~ gauge}~(\partial^i A_i \approx 0)~:~ \int d^2 z c_{j}
  ({\bf x}-{\bf z}) A^j({\bf z}) =0,~
\mbox{b)~Axial ~gauge}~(A_1 \approx 0)~:~ \partial^2_z c_2({\bf x}-{\bf z})
  =-\delta^2 ({\bf x}-{\bf z}),~ 
\mbox{c)~Weyl ~gauge}~(A_0 \approx 0)~:~ \partial^j_z c_j({\bf x}-{\bf z})= 
 -\delta^2 ({\bf x}-{\bf z})$. These results are generally valid for
any other gauge theories when they are formulated by our gauge
invariant formulation. 
Note that these results are different from recent claims of Refs. [2,
3] except in the  case of Coulomb gauge [15]. 
Moreover, the Weyl gauge does not modify the equation 
for $c_k$ from (4).
Then, using these relations and (8), we could consider the gauge fixed 
results directly from GIF. However, as can be observed in these
examples, gauge fixings restrict the solution space in
general. Therefore, all the quantities which appear in (6) are gauge
invariant for each solution hypersurface which is selected by gauge
fixing, but their functional form may be different depending on the
gauges. 
Here we show the case of Coulomb gauge, which is 
found to have a special meaning. In this case we find, using 
$c^j({\bf x}-{\bf z}) =-(1/2 {\pi})
 ({\bf x}-{\bf z})^j/|{\bf x}-{\bf z}|^2$ which solves the  equation
in ``a)'',   
\begin{eqnarray}
\Xi^{12}({\bf x})=\frac{1}{2 \pi \kappa} Q,~~~ \Xi^{0i}({\bf x})=
\frac{1}{2 \pi \kappa} \int d^2 z \frac{({\bf x} -{\bf z})^i
 ({\bf x}-{\bf z})^k}{|{\bf x}-{\bf z}|^2} \epsilon_{jk} J^j({\bf z}),
\end{eqnarray}
where $Q=\int d^2 z J_0$ [16].
This is exactly Hagen's rotational anomaly and Coulomb gauge restoring term in
 the Lorentz transformation, respectively [6]. 
Furthermore, the basic brackets defined in (7) is found to be the 
usual Dirac brackets in the Coulomb gauge by noting 
$\Delta({\bf x}-{\bf y})=0$ and 
$\xi_{ij} ({\bf x}-{\bf y})=-\epsilon_{ij} \delta^2({\bf x}-{\bf y})$ in this 
case. 
[Upon using $c^j({\bf x}-{\bf z})=\partial^j_{z} \frac{1}{2 \pi} 
ln |{\bf x}-{\bf z}|$ and performing the integration by parts, we obtain 
$\Delta({\bf x}-{\bf y})=-\frac{1}{2 \pi} \oint _{S^1_{R \rightarrow \infty}}
 d \theta \hat{\theta} \cdot \hat{r} ln R =0$,  where the integration is 
evaluated on a circle with infinite radius $R$, polar angle $\theta$, and 
their corresponding (orthogonal) unit vectors $\hat{r},~\hat{\theta}$. 
Moreover, using the antisymmetry 
$c^j({\bf x}-{\bf z})=-c^j({\bf z}-{\bf x})$ and (4), the expression for 
$\xi_{ij}$ given above can be verified.]
This implies that the gauge invariant operator $\hat{\phi}_{op}$ satisfies 
the boson commutation relation, $[ \hat{\phi}_{op}({\bf x}),\hat{\phi}_{op}
({\bf y})]=0$ in this case.
Here, we note the special importance of the Coulomb gauge in that the original
 fields $\phi,~ \phi ^*, ~A_{\mu}$ themselves are already gauge invariant 
fields 
such that they already have the full anomaly structures of (6). Furthermore, 
this gauge is the simplest one to obtain the anomalous spin of the original 
matter field $\phi$ as $\Xi^{12}$ of (9) since this does not have other gauge
 restoring terms as in the rotationally non-symmetric 
gauge. This is made clear by noting the relation [5,6]
$M^{12}_s \approx M^{12}_c -({\kappa}/{2}) 
\int d^2 z \partial ^k (z^k A^l
A^l-z^l A^l A^k )$, 
where $M^{12}_c$ is the canonical angular momentum
$M^{12}_c =\int d^2 z \left\{ \epsilon_{lk} z^l [\pi \partial^k \phi + 
(\partial^k \phi)^* \pi^* ] -\kappa z^l A^l (\partial^k A^k) +({\kappa}/{2})
 \partial^k (z^l A^l A^k)\right\}$.
The surface terms in $M^{12}_s -M^{12}_c$ and $M^{12}_c$, which are
gauge invariant for the rapidly  
decreasing gauge transformation function $\Lambda$, give the gauge
independent spin terms  
``$({1}/{4 \pi \kappa}) Q^2$" [6-8] (unconventional) and ``$0$" (conventional)
 in $M^{12}_s$, respectively. 
[Explicit manipulations of the gauge independence of the unconventional term
have been established only for limited class of gauges [6-8]. But these 
results are generalized to the case of general gauges due to the gauge 
invariance of the term.]
From which the anomalous spin $\frac{Q}{2 \pi \kappa}$ of (9) for the
matter field is  
readily seen to follow for general gauges [17].
On the other hand, the second term of $M^{12}_c$, which vanishes only in the 
Coulomb gauge, gives for the general gauges 
the gauge restoring contribution to the rotation transformation for the matter
field.

Before completing our analysis we note that the canonical (Noether) 
Poincar\'e generators can not be considered as the physical ones since the 
canonical boost generators $M^{0i}_c \approx M^{0i}_s +\frac{\kappa}{2} 
\int d^2 z A^0 \epsilon_{ij} A^j$ are not gauge invariant due to the last
gauge variant term ``even on the constraint surface'' and do not
satisfy the (classical) 
Poincar\'e algebra:
\begin{eqnarray}
\{{M}^{0i}_{c}, {M}^{12}_{c}\} &\approx&
 -\epsilon_{ij} {M} ^{0j}_{c}+\frac{\kappa}{2} \epsilon_{ij} \int d^{2}z 
\partial^{l} (\epsilon_{kl}z^{k}A^{0} {A}^{j}), \nonumber \\
\{{M}^{0i}_{c}, {M}^{0j}_{c} \} &\approx&
  -\epsilon_{ij} {M}^{12}_{c}- \epsilon_{ij}\frac{1}{4 \pi \kappa} Q^2 
 +\epsilon_{ij} \frac{\kappa}{2} \int d^2 z \left[\frac{5}{2} A^{2}_{0} 
+{A}^{k} {A}^{k} +\partial^{0}(z^{k} A^{0}{A}^{k}) \right].
\end{eqnarray}
It is the improved generators (5), constructed from the symmetric 
energy-momentum tensor, which are (manifestly) gauge invariant and
obey the classical Poincar\'e algebra. 
Hence the improved generators (5) have a unique meaning consistently with 
Einstein's theory of gravity [18]. 
From this fact it is seen that the anomalous spin of the relativistic matter, 
which comes only from $M^{12}_s$, is not artificial, contrary to recent claim 
of Graziano and Rothe [6].  Furthermore, this uniqueness of anomalous spin is 
in contrast to the anomalous statistics, which has only artificial meaning in 
this case. This is because we can obtain in any field theories [13] any 
$arbitrary$ statistics by constructing gauge invariant exotic operator of the 
form of Semenoff and its several variations [6]. In this sense the 
relativistic CS gauge theory does not respect the $spin-statistics$ 
relation [12] in agreement with Hagen's result [6, 13].
Here, we add that the situation of non-relativistic CS gauge theory is not 
better than this relativistic case. This is because even though the anomalous 
statistics is uniquely defined by removing the gauge field (in this case the 
gauge field is pure gauge due to point nature of the sources in 
non-relativistic quantum field theory) the anomalous spin has no unique 
meaning [6, 7].

In summary, we have considered a new GIF consistent with GFF. 
Our formalism is new in the following three points. (A) We introduced the 
assumption that there be no translation transformation anomaly for gauge 
invariant quantities ${\cal F}_{\alpha}$. From this assumption, we obtained 
several new conditions for the dressing function $c_{k}({\bf x}, {\bf z})$, 
which are crucial in our development. (B) We introduced the master formula 
(8), which allowed
matching to the gauge fixed system. (C) We found and used the manner how the 
equation of the dressing function $c_{k}({\bf x}, {\bf z})$ are modified
after gauge fixing. Using this formulation, we have obtained a novel GIF, 
which is consistent with the conventional GFF: The former formulation provides
 exactly the rotational anomaly of the latter.
Hence, in our formulation there is no inconsistency, as in the previous gauge 
independent formulation of Ref. [5].
As a byproduct, we explicitly found that the anomalous spin of the charged 
matter has a unique meaning [6]. This is due to the uniqueness of the 
Poincar\'e generators when constructed from the symmetric energy-momentum 
tensor.

We would like to 
conclude with three additional comments. First, in our formulation, there is 
no gauge non-invariance problem of Poincar\'e generators on the physical 
states.
This is  
essentially due to absence of additional terms proportional to constraints in 
the generators of (5), in contrast to the old formulation of Dirac [1]. 
Second, the master formula (8), which guarantees the classical
Poincar\'e covariance  
of our CS gauge theory in all gauges, also works in all other gauge theories. 
Hence as far as the $gauge~ dependent$ operator ordering problem does not 
occur, the $quantum$ Poincar\'e covariance 
for one gauge guarantees also the covariance for all other gauges. The gauge 
independent proof of quantum covariance has been an old issue in quantum 
field theory and now it is reduced to the solvability of the problem of the 
gauge dependent operator ordering. Finally, it has been recently reported 
that canonical 
Poincar\'e generators in QED or QCD also do not satisfy the Poincar\'e 
algebra. But it's origin is different from ours [2].\\

One of us (M.-I.Park) would like to thank Prof. R. Jackiw for valuable 
conversations and acknowledge the financial support of Korea Research 
Foundation made in the program year 1997. The present work was supported in 
part by the Basic Science Research Institute Program, Ministry of Education, 
Project No. BSRI-97-2414, and in part by the U.S. Department of Energy
(D.O.E) under cooperative research  
agreement No. DF-FC02-94ER40818.

\newpage
\begin{center}
{\large \bf References}
\end{center}
\begin{description}

\item{[1]} P. A. M. Dirac, Can. J. Phys. {\bf 33}, 650 (1955).

\item{[2]} M. Lavelle and D. McMullan, Phys. Lett. {\bf B312}, 211 (1993);  
{\bf 329}, 68 (1994); Phys. Rep. {\bf 279}, 1 (1997).

\item{[3]} P. Gaete, Z. Phys. {\bf C76}, 355 (1997); T. Kashiwa and N.
 Tanimura, Phys. Rev. {\bf D56}, 2281 (1997).

\item{[4]} P. A. M. Dirac, {\it Lecture on Quantum Mechanics} (Belfer Graduate
School of Science, Yeshiva University Press, New York, 1964).

\item{[5]} R. Banerjee, Phys. Rev. Lett. {\bf 69}, 17 (1992);
Phys. Rev. {\bf D48}, 2905 (1993);
R. Banerjee and A. Chatterjee, Ann. Phys. (N.Y.) {\bf 247}, 188 (1996). 

\item{[6]} C. R. Hagen, Ann. Phys. (N.Y.) {\bf 157}, 342 (1984);
Phys. Rev. {\bf D31}, 2135 (1985); G. W. Semenoff,
Phys. Rev. Lett. {\bf 61}, 517 (1988); R. Jackiw and S. Y. Pi,
Phys. Rev. {\bf D42}, 3500 (1990); 
E. Graziano and K. D. Rothe, {\it ibid.}, {\bf D49}, 5512 (1994).

\item{[7]} C. R. Hagen, Phys. Rev. {\bf D31}, 331 (1985); {\bf D31}, 848
 (1985).

\item{[8]} H. Shin, W.-T. Kim, J.-K. Kim, and Y.-J. Park, Phys. Rev. 
{\bf D46, } 2730, (1992).

\item{[9]} F. J. Belinfante, Physica (Utrechet) {\bf 7}, 449 (1940).

\item{[10]} L. Faddeev and R. Jackiw, Phys. Rev. Lett. {\bf 60}, 1692 (1988).

\item{[11]} C. G. Callan, S. Colman, and R. Jackiw, Ann. Phys. {\bf 59}, 42
 (1970).

\item{[12]} F. Wilczek, Phys. Rev. Lett. {\bf 48}, 1144 (1982); {\bf 49}, 957 
(1982).

\item{[13]} C. R. Hagen, Phys. Rev. Lett. {\bf 63}, 1025 (1989); {\bf 70}, 
3518 (1993).

\item{[14]} I. A. Batalin and I. V. Tyutin, Int. J. Mod. Phys. {\bf A6}, 3245
 (1991); Y.-W. Kim, M.-I. Park, Y.-J. Park, and S.-J. Yoon, {\it ibid.},
 {\bf A12}, 4217 (1997).

\item{[15]} Authors of Refs. [2, 3] considered $\int d^2 z c_j ({\bf
x}-{\bf z} ) A^j ({\bf z}) =0$ even ``b)'' and ``c)'' cases. But then,
the manifestly gauge invariant fields in (3) are not gauge invariant
under the residual gauge symmetries $ \phi \rightarrow e^{-i\Lambda }
\phi,~A_{\mu} \rightarrow A_{\mu} +\partial_{\mu} \Lambda$ with $x^1$
and $x^0$ independent $\Lambda$ for ``b)'' and ``c)'', respectively. 
 
\item{[16]} In the Weyl gauge, the simplest solution is $c_1=0,~
c_2({\bf x})=\delta(x^1) \epsilon (x^2)$ with a step function
$\epsilon (x)$. This corresponds to a different solution hypersurface
to the Coulomb gauge and hence it's related anomalous terms in (6)
have different functional form to (9) even though they are gauge
invareinat on it's own hypersurface: $\Xi^{12} =\frac{1}{\kappa} \int
^{\infty}_{-\infty} d y^2 J_0 (x^1, y^2),~ \Xi^{02} =\frac{1}{\kappa}
\int ^{\infty}_{-\infty} d y^2 J^1 (x^1, y^2),~ \Xi^{01} =0$. 

\item{[17]} This is because the commutation relation $[ Q, \phi (x) ]=
 \phi (x)$ is gauge independent in the general gauges  
 $\int d^2 z K_{\mu}({\bf x}, {\bf z}) A^{\mu}({\bf z}) \approx 0$ with kernel
 $K_{\mu}({\bf x},{\bf z})$.
%\item{[18]} The preferred property of the symmetric energy-momentum tensor 
%compared to
%canonical one by the gauge invariance was examined by S. Deser and J. G. 
%MaCarthy, Nucl. Phys. {\bf B344}, 763 (1990). But they missed the important 
%role of the
%gauge invariance, which is genuine for the CS gauge theory, 
%for the Poincar\'e generators, which are spatial integration of the 
%energy-momentum tensor and its multiplication by the field point $x^{\mu}$.
\item{[18]} There may be other differently improved generators depending on 
what gravity theory is chosen like as in Ref. [11]. But  we do not consider 
this possibility in this paper.
\end{description}
\end{document}